\documentclass[letterpaper]{article} 
\usepackage{aaai2026}  
\usepackage{times}  
\usepackage{helvet}  
\usepackage{courier}  
\usepackage[hyphens]{url}  
\usepackage{graphicx} 
\usepackage{natbib}  
\usepackage{caption} 
\usepackage{transparent}
\usepackage{amsmath}
\usepackage{amsfonts}
\usepackage{siunitx}
\usepackage{tabularx}
\DeclareUnicodeCharacter{2212}{-}

\nocopyright

\title{Modelling the Effects of Hearing Loss on Neural Coding in the Auditory Midbrain with Variational Conditioning}
\author {
    Lloyd Pellatt\textsuperscript{\rm 1},
    Fotios Drakopoulos\textsuperscript{\rm 1},
    Shievanie Sabesan\textsuperscript{\rm 1},
    Nicholas A. Lesica\textsuperscript{\rm 1,2}
}
\affiliations {
    \textsuperscript{\rm 1}UCL Ear Institute, London, UK \\
    \textsuperscript{\rm 2}Perceptual Technologies, London, UK\\
    l.pellatt@ucl.ac.uk, f.drakopoulos@ucl.ac.uk, s.sabesan@ucl.ac.uk, n.lesica@ucl.ac.uk
}

\begin{document}

\maketitle

\begin{abstract}
The mapping from sound to neural activity that underlies hearing is highly non-linear. The first few stages of this mapping in the cochlea have been modelled successfully, initially with biophysical models built by hand and, more recently, with DNN models trained on datasets simulated by these models. Modelling the auditory brain has been a challenge because central auditory processing is too complex for models to be built by hand, and datasets for training DNN models directly have not been available. Recent work has taken advantage of large-scale high resolution neural recordings from the auditory midbrain to build a DNN model of normal hearing with great success. But this model assumes that auditory processing is the same in all brains, and therefore it cannot capture the widely varying effects of hearing loss.  
     
We propose a novel variational-conditional model to learn to encode the space of hearing loss directly from recordings of neural activity in the auditory midbrain of normal and noise exposed animals. With hearing loss parametrised by only 6 free parameters per animal, our model accurately predicts 62\% of the explainable variance in neural responses from normal hearing animals and 68\% for hearing impaired animals, comparable to state of the art animal specific models. We demonstrate that our model can be used to simulate realistic activity from out of sample animals by fitting only the learned conditioning parameters with Bayesian optimisation, achieving crossentropy loss within 2\% of the optimum in 15-30 iterations. Including more animals in the training data improved the performance on unseen animals. This model will enable future development of parametrised hearing loss compensation models trained to directly restore normal neural coding in hearing impaired brains, which can be quickly fitted for a new user by human in the loop optimisation.
\end{abstract}


\section{Introduction}
\label{sec:Introduction}

Computational models of neural responses to sound are widely used in the study of hearing and hearing loss \cite{auditorymodels}. Many current models focus on predicting responses in the cochlea and auditory nerve. These models can be parametrised to simulate different types of hearing loss, drawing on known biophysical causes of hearing loss such as inner/outer hair cell loss or cochlear synaptopathy \cite{verhulst2018computational}, but with important limitations. The engineered nature of these models means that they make assumptions about the effects of hearing loss, and they are slow to compute and non-differentiable. In recent years deep learning has been used to approximate the non-linear mapping from sound to neural activity expressed by these analytically derived models in a more efficient and differentiable way \cite{baby2021convolutional, nagathil2023wavenet, peterleerconditional, drakopoulos2023neural}. But these models are trained on data generated by the original models, thereby inheriting their biases and assumptions. Even if these models were to provide perfect predictions of activity in the healthy and damaged cochlea, there are other important changes in the central auditory pathway due to brain plasticity that would not be accounted for \cite{nelson2004relationship, auerbach2014central}. 

Recently, a unique large-scale dataset of neural activity in the inferior colliculus (IC) of a large number of normal hearing (NH) and noise exposed gerbils has been collected by Sabesan et al. \cite{10.7554/eLife.85108}. This has enabled the development of DNN models trained on real neural activity \cite{10.7554/eLife.85108, drakopoulos2025modelling}. These models are fast, differentiable and can incorporate all of the complex, non-linear properties of auditory processing up to the level of the midbrain. The limitation of this more direct approach is that it can only be used to simulate brain activity for specific animals from which there are existing recordings. While the models have been shown to accurately simulate both hearing impaired (HI) and NH brain activity \cite{10.7554/eLife.85108}, this requires training a new model from scratch for every brain. \citeauthor{drakopoulos2025modelling} have shown that multi-branch DNNs trained to predict neural activity in the IC of multiple NH animals can capture shared dynamics across different brains \cite{drakopoulos2025modelling}. However, while every healthy auditory system can be assumed to exhibit similar patterns of neural activity, each impaired system will exhibit unique distortions \cite{parida2022underlying}.

Drawing on conditional and `attribute-based' generative models \cite{kawai2020attributes,NEURIPS2023_fe51de4e,kapoor2024latent,azabou2023unified}, we propose a complete model of healthy and impaired neural coding in the auditory midbrain parametrised by a small set of parameters $\psi$ which are directly learned from data and which represent a low dimensional encoding of the full spectrum of hearing loss. We show that the model learns to simulate neural coding in a wide variety of hearing impaired and normal hearing brains with accuracy comparable to current state of the art model ICNet \cite{drakopoulos2025modelling}, without explicit assumptions as to the effects of hearing loss. We demonstrate that the space of conditioning parameters learned by the model is smooth and that the conditioning parameters can be optimised independently of the rest of the model weights to predict neural activity in animals on which it was not trained. This constitutes the first model of the auditory midbrain which can simulate both healthy and impaired neural activity across a wide spectrum of hearing loss.

\section{Related Work}
\label{sec:RelatedWork}

The ICNet model proposed by \citeauthor{drakopoulos2025modelling} has been shown to accurately simulate neural coding in the IC, in the form of multi-unit activity (MUA) on each of 512 recording channels in 1.3 ms time bins. ICNet follows an encoder-decoder structure, where the convolutional encoder is used to extract salient features from an input sound $s$ which are then interpreted by a pointwise 1-D convolutional decoder to predict a five class categorical distribution over the number of spikes recorded in each time bin. The model then samples from this distribution to predict the neural activity $\hat{R} \sim f(s | \theta)$, where $\theta$ represents the model weights.

This model architecture has been shown to reproduce the real MUA recorded from many individual animals with both normal hearing status and varying degrees of hearing loss with remarkable accuracy \cite{10.7554/eLife.85108,drakopoulos2025modelling}. A multi-branch architecture was shown to better model the neural responses of up to 9 NH animals, using a shared sound encoder network with a 64 channel bottleneck to capture the shared latent dynamics and unique decoders for each animal \cite{drakopoulos2025modelling}. 

\subsection{Conditional models}
\label{sec:ConditionalModels}

\citeauthor{NEURIPS2023_fe51de4e} propose a `neural data transformer' architecture which learns a joint embedding of neural activity in the motor cortex recorded by brain-computer interfaces in monkeys and humans \cite{NEURIPS2023_fe51de4e}. The model is given conditioning variables corresponding to subject and session IDs, and learns to reconstruct neural activity and predict associated behaviours. \citeauthor{azabou2023unified} use a GPT architecture to model neural activity in motor cortical regions of non-human primates, and decode behaviours from the neural code \cite{azabou2023unified}. \citeauthor{kapoor2024latent} introduce a latent diffusion based model for conditional generation of neural spike data recorded from the motor cortex of non-human primates while performing reach gestures, conditioned on the reach direction or trajectory \cite{kapoor2024latent}. All three of these models take an auto-encoder approach where learned conditional embeddings are used to help reconstruct the neural activity. We apply a similar concept to individualise the latent feature representations of ICNet, while keeping a simple shared decoder to ensure that the conditioning mechanism learns only to encode differences due to hearing loss.

\citeauthor{peterleerconditional} propose a conditional framework in which weights for a DNN-based model of the cochlea are generated from an audiogram (a standard audiometric measure of hearing loss) by a weight generating network \cite{peterleerconditional}. This approach differs from our work in two important ways. First, this model includes only the most peripheral stages of the auditory system and functional effects of hearing loss on these stages, whereas ours reflects the full complexity of normal and impaired processing that takes place in the cochlea, brainstem and midbrain. Additionally, our method learns the underlying structure of the space of hearing loss directly from the distribution of hearing impaired brains in the training dataset, rather than relying on audiometric thresholds. This is advantageous since audiometric thresholds have been shown to be inadequate descriptors of hearing loss, as clearly illustrated by the prevalence of `hidden hearing loss', a condition in which a patient exhibits impaired perception of complex sounds despite normal audiometric test results \cite{plack2014perceptual}. Any perceptual difficulties which are not reflected in audiometric measurements should nonetheless result in distorted neural activity in the IC \cite{parida2022underlying}.

\section{Variational-Conditional Model}
\label{sec:ModelArchitecture}

We propose an extension of ICNet to account for hearing loss by assuming that the effects of hearing loss on the neural code can be parametrised by some small set of parameters $\psi$, so that the model should approximate a function $f(s | \theta, \psi)$ where $\theta$ is generic while $\psi$ encodes the individual effects of hearing loss on the latent features of the sound. Our objective is to compress the effects of hearing loss into a latent space $\psi$ with as few dimensions as possible while maintaining comparable prediction accuracy to ICNet.

\graphicspath{{./figures/psi_arch}}

\begin{figure*}[htbp]
\centering
\begin{minipage}{0.95\linewidth}
\def\svgwidth{0.95\linewidth}
\fontsize{9pt}{9pt}\selectfont
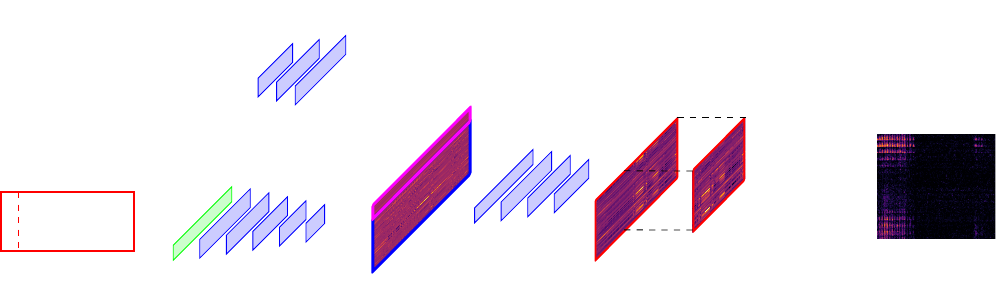
\end{minipage}%
\caption{Architecture of $\psi$-ICNet. An input sound is processed by a convolutional encoder, consisting of a SincNet layer with 48 filters followed by 5 causal convolutional layers with 246 kernels of size 60 and a final layer with 64 kernels. The resulting representation $\hat{r}_b$ encodes latent dynamics shared across animals. The variational conditioning module introduces animal-specific information in the form of a set of values $\psi$ as described in figure \ref{fig:time_degradation_module}. A three layer convolutional network with $N_\psi$, $2N_\psi$ and $3N_\psi$ kernels (where $N_\psi$ is the number of $\psi$ parameters) expands $\psi$ into a feature map $\hat{r}_{\psi}$ of size $320 \times 3N_{\psi}$, which is concatenated with $\hat{r}_b$. A further 4 convolutional layers containing progressively fewer kernels of size 16 then combine the shared and animal-specific features, reducing the dimensionality back to $320 \times 64$. This produces an individualised representation $\hat{r}_b'$, which is cropped to remove context. The decoder is shared between animals and outputs a categorical probability distribution $p(\hat{R}|s, \psi)$ over the number of spikes at each channel and timestep, from which we sample to predict $\hat{R}$.}
\label{fig:psinet_arch}
\end{figure*}

\begin{figure}[htb]
\centering
\begin{minipage}{\linewidth}
\def\svgwidth{\textwidth}
\fontsize{9pt}{9pt}\selectfont
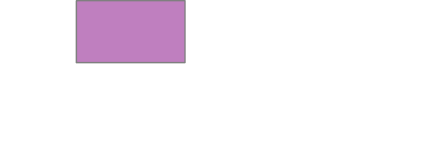
\end{minipage}%
\caption{Architecture of the variational conditioning module which encodes the effects of hearing loss on the generic latent representation. $\mu_{\psi}^t$ is a set of learned weights which form the means of a multivariate Gaussian distribution from which we sample (using the reparameterisation trick)  to produce values which modify the bottleneck. The means are modified by the time transfer function which accounts for the change over time of the state of the animal during recording. A sigmoid function scales the modified weights between zero and one, and we sample from a normal distribution with learned covariance matrix $\Sigma_{\psi}$, which is diagonal.}
\label{fig:time_degradation_module}
\end{figure}

Figure \ref{fig:psinet_arch} shows the architecture of our conditional model. We use the same sound encoder as ICNet to extract a 64 channel feature map. This shared representation is then individualised by incorporating a set of learned conditioning parameters $\psi$ which encode the hearing related differences between brains. The conditional parameters $\psi$ are the only part of the model which is learned uniquely for a given animal \textemdash all other parameters in the model are shared between animals. The means $\mu_{\psi}^t$ are dependent on the measurement time of the MUA according to the time transfer function described below, to account for the non-stationarity of the neural responses over time \cite{drakopoulos2025modelling}. Once the latent feature maps have been conditioned, we pass the individual representations $\hat{r}_b'$ through a shared decoder which projects the latent features to an MUA representation.

\subsection{ABRNet}
\label{sec:ABRNet}

To test our hypothesis that learning conditioning parameters directly from the data can produce a better conditioning space than an equivalent approach using audiometric measurements, we implemented a variation of our model architecture where the learned means $\mu_{\psi}$ of the conditioning variables are replaced by the auditory brainstem response (ABR) threshold measurements from each animal in response to tones at 1, 2, 4, 8, and 16 kHz. These ABR thresholds constitute a description of an animal's hearing loss profile comparable to the audiogram of a human listener, and therefore ABRNet should be able to learn to generate individualised responses to the extent that the audiogram captures the relevant information. We did not add Gaussian noise to the ABR thresholds. The model is otherwise identical to $\psi$-ICNet.

\subsection{Modelling changes in neural responses over time}
\label{sec:TimeDegradation}

To better model the neural activity over the whole duration of the recording and to capture elements of the degradation over time of each animal which mimic an increase in the severity of hearing loss, we modify the mean of the learned $\psi$ distributions for each animal as shown in figure \ref{fig:time_degradation_module}, proportionally to the time elapsed since the beginning of the recording. We model the degradation of the neural activity of a given animal according to the following observations:

\begin{itemize}
\item It should be monotonic with respect to time.
\item It should be negligible for the first hour of the recording.
\item It should either increase quickly and saturate, or increase slowly and accelerate until the end of the recording.
\end{itemize}

We use a piecewise transfer function which modifies each $\psi$ parameter according to the time $t$ (in hours divided by 10) elapsed since the beginning of the recording as follows:

\footnotesize
\begin{equation}
T_{\psi}(\psi, t) = \begin{cases}
\psi, & \text{if } t \leq o_t \\
\psi + \vec{s}_t\odot\sigma(\vec{W}_t(t-\vec{o}_t)-6)-I, & \text{otherwise} 
\end{cases}
\end{equation}
\normalsize

where $\odot$ denotes an element-wise product, $\vec{s}_t$ is a set of trainable parameters determining the maximum extent of the time degradation, $\vec{W}_t$ determines the slope of the degradation, and $\vec{o}_t$ is the `onset' which determines when the time degradation begins. $\sigma$ is the sigmoid function $\sigma(x) = \frac{1}{1+e^{-x}}$. $I = \vec{s}_t \odot \sigma(-6)$ is the `intercept' which must be subtracted to avoid a discontinuity at $t=o_t$ \textemdash the fixed bias of -6 ensures the rate of increase is close to zero when $t=o_t$.

Values of $\vec{s}_t$ are constrained between 0 and 20, keeping the scale of the time effects equal to the original range of $\psi$ (between -10 and 10). Values of $\vec{W}_t$ are constrained between between 0 and 50, and $\vec{o}_t$ is constrained between 0.05 and 0.5 \textemdash i.e. between 30 minutes and 5 hours into the recording. 

\subsection{Loss function and hyperparameters}
\label{sec:LossFunction}

We trained each model with a loss function specified by:

\begin{equation}
\mathcal{L} = \mathcal{L}_{CE}(R,p(\hat{R})) + \alpha_{KL} D_{KL}+ \alpha_{ABR} \mathcal{L}_{ABR},
\end{equation}

$\mathcal{L}_{CE}$ is the crossentropy loss between the model output $p(\hat{R})$ and a one-hot representation of the target MUA $R$, $D_{KL}$ is the KL divergence $KL(\mathcal{N}(\mu^t_{\psi}, \Sigma_{\psi})|| \mathcal{N}(\mu^t_{\psi}, \Sigma^*))$ between the $\psi$ distributions and a target distribution with covariance matrix $\Sigma^* = \frac{1}{N_A}\cdot\mathbb{I}$ where $N_A$ is the number of animals, and $\alpha_{KL}$ is a weighting coefficient set to $10^{-3}$.

To further encourage clustering of animals with similar hearing loss in the $\psi$ space, we applied a small penalty $\mathcal{L}_{ABR}$ in the first 10 epochs for encoding animals with dissimilar ABR thresholds close together, given by

\begin{align}
    L_{ABR} &= \sum_{i \neq j} A_{ij} e^{(-10 \frac{d_{ij}}{\bar{d}})},
\end{align}

where $d_{ij} = \frac{1}{n}\sum_{k=1}^n |\mu_{\psi,ik} - \mu_{\psi,jk}|$ is the mean absolute difference between learned means for animals $i$ and $j$ and $A_{ij}  = \sqrt{\frac{1}{m}\sum_{k=1}^m (T_{ik} - T_{jk})^2}$ is the RMSE between ABR thresholds $T$ of animals $i$ and $j$, where \(m\) is the number of frequencies at which ABR thresholds were measured. We scale the distances by $\bar{d} = \sqrt{\frac{n}{6} - \frac{7}{120}}$, which gives the expected distance between two random points in a unit hypercube of dimension $n$, to ensure that the penalty is applied consistently for different sizes of $\psi$. We used a weighting of $\alpha_{ABR}=0.02$ for the first 10 epochs and 0 afterwards.

We trained each model using the Adam optimiser \cite{Adam} for up to 120 epochs, stopping early if validation loss did not improve for 10 successive epochs. We set the learning rate $\alpha$ to $4e{-4}$ and reduced it by half if the validation loss did not improve for 4 successive epochs. The encoders for all models were initialised with weights from an ICNet model trained on 12 NH animals. 

\subsection{Performance metrics}
\label{sec:PerformanceMetrics}

To assess how well our models capture the repeatable signal in the noisy recorded data we take two recordings $R_1$ and $R_2$ of MUA produced from identical sounds, we predict two responses $\hat{R}_1$ and $\hat{R}_2$ to the same sound, and compute the fraction of explainable variance explained (FEVE), weighted by the cross-trial covariance of each channel, given by:

\footnotesize
\begin{equation}
F_W = \frac{1}{M}\sum_{i=1}^{M} w_i \left(1 - \frac{\frac{1}{2}\left(C(R_1^i, \hat{R}_1^i) + C(R_2^i, \hat{R}_2^i) \right) - \sigma^2_{i}}{\frac{1}{2}(Var(R_1^{i}) + Var(R_2^{i})) - \sigma^2_{i}}\right),
\label{eq:FEVE}
\end{equation}
\normalsize
where $C(X,Y) = \frac{1}{T}\sum\limits_{t=1}^{T}(X_t - \mathbb{E}_C[Y_t])^2$ and $w_i = Cov(R_1^{i}, R_2^{i})$ is the cross-trial covariance of channel $i$, $\mathbb{E}_C$ is the expectation over spike counts of the model output \textemdash $p(\hat{R}|s,\psi)$ for $\psi$-ICNet, $\sigma^2_{i} = \frac{1}{2}Var(R_1^{i} - R_2^{i})$ is a measure of the noise floor in the MUA for channel i, and $M$ and $T$ are the size of the channel and time dimension. 

We also computed the covariance weighted KL divergence between the normalised spike count histograms $H$ of the true and predicted MUA, given by:

\begin{equation}
KL_W = \sum_{i=1}^{M} w_i \cdot \frac{1}{2}\left(KL(\hat{H}_1^i || H_1^i) + KL(\hat{H}_2^i || H_2^i)\right),
\end{equation}
where $H^i_j$ is computed over all timesteps in recording $j$ for channel $i$ and $\hat{H}$ is the histogram of predicted spikes.

\subsection{Computational resources}
\label{sec:ComputationalResources}

Each training run of the multi-animal models described in this paper required around 10-20 GPU hours of training time, while the single animal models required 3-5 GPU hours to train. All experiments were performed locally on a GPU server with 4 RTX 4090 GPUs with 24GB VRAM and 256GB of system memory. In total, experiments reported in this paper used around 300-500 GPU hours of compute. We estimate that the research so far has consumed several thousand GPU hours including preliminary work and other experiments whose results are not reported in this paper.

\section{Dataset Collection and Preparation}
\label{sec:DatasetCollectionandPreparation}

We trained our models on MUA recorded from 9 gerbils, 3 of which had normal hearing and 6 of which were noise exposed with symmetric hearing impairment \textemdash determined by comparing ABR thresholds across ears, according to the procedure outlined in \cite{10.7554/eLife.85108}. The dataset for each animal consists of 512 channel MUA, where each channel represents the spiking activity of a small group of neurons in the IC. Audio was presented at a sampling frequency of 24414.06 Hz. MUA was extracted from neural activity recorded at 20 KHz and resampled to 762.94 Hz (temporal precision of 1.3 ms). To prepare the data, we segmented the audio and MUA into frames of 8192 samples (0.33 seconds). We included context of 2048 samples (0.08 seconds) at the beginning of each frame of audio, which was cropped from the latent representation as shown in figure \ref{fig:psinet_arch}. 

The training dataset contained 7.8hrs of data, consisting of a mixture of speech, processed speech, speech in noise, music and dynamic ripples \cite{drakopoulos2025modelling}, split 90:10 between training and validation. We tested the models using a variety of 30s segments of sounds from the same classes but from different sources (e.g. speech corpora). 

\subsection{Alignment between brains}
\label{sec:Alignment}

Even two animals with normal hearing will produce subtly different neural responses, due to differences in the spatial organisation of information in the IC. To ensure that the parameter space learned by our model primarily encodes differences related to hearing loss, we must remove this source of variability from the dataset before training the model. It is critical that any alignment method retains differences between animals which are due to hearing loss (including differences of scale), while minimising differences due to spatial organisation. This makes most common methods of aligning neural activity between brains unsuitable.

To accomplish this, we used optimal transport to find a permutation of the channels of each brain recording that minimised the Wasserstein distance to a common NH reference animal. This ensures that no scale information is lost after alignment by only allowing shuffling of channel order. Alternative methods suitable for this problem include spatial transforms (including Procrustes analysis \cite{Procrustes} and spatial transformer networks \cite{jaderberg2015spatial}) and maximum covariance analysis. We have not observed significant differences between these approaches. 

\section{Comparison Against ICNet}
\label{sec:ComparisontoSingleBranchModel}

We trained $\psi$-ICNet, ABRNet and individual ICNets on 9 animals (3 NH and 6 HI) according to the protocol described above, and evaluated their performance on our testing dataset of two-trial sounds. For $\psi$-ICNet we trained models with 3 and 6 $\psi$ parameters. Increasing the number of parameters did not improve performance but did result in models which took significantly longer to fit to unseen animals (not reported).

\subsection{Results}

\begin{figure*}
\begin{minipage}[b]{0.5\textwidth}
\centering 
\def\svgwidth{\textwidth}
\fontsize{7pt}{7pt}\selectfont
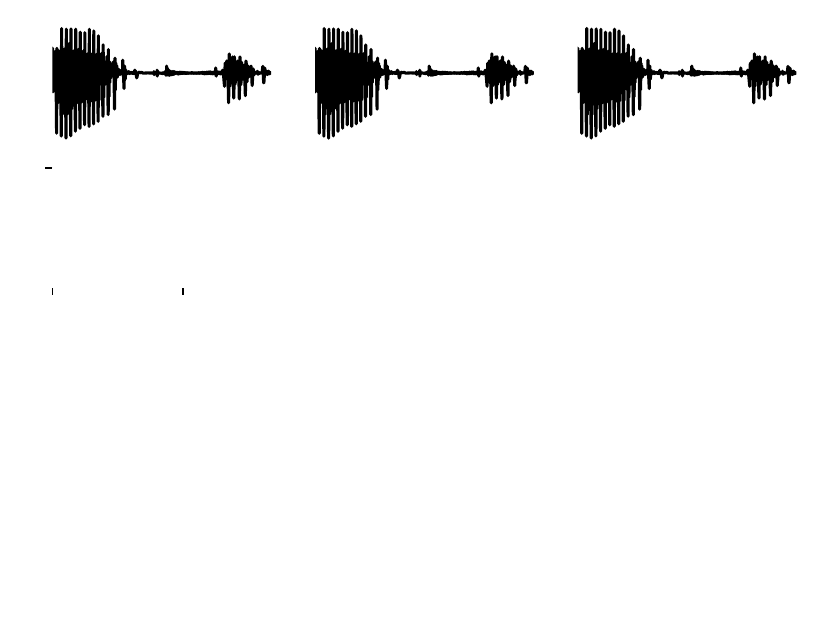\\
\end{minipage}%
\begin{minipage}[b]{0.5\textwidth}
\centering
\def\svgwidth{\textwidth}
\fontsize{7pt}{7pt}\selectfont
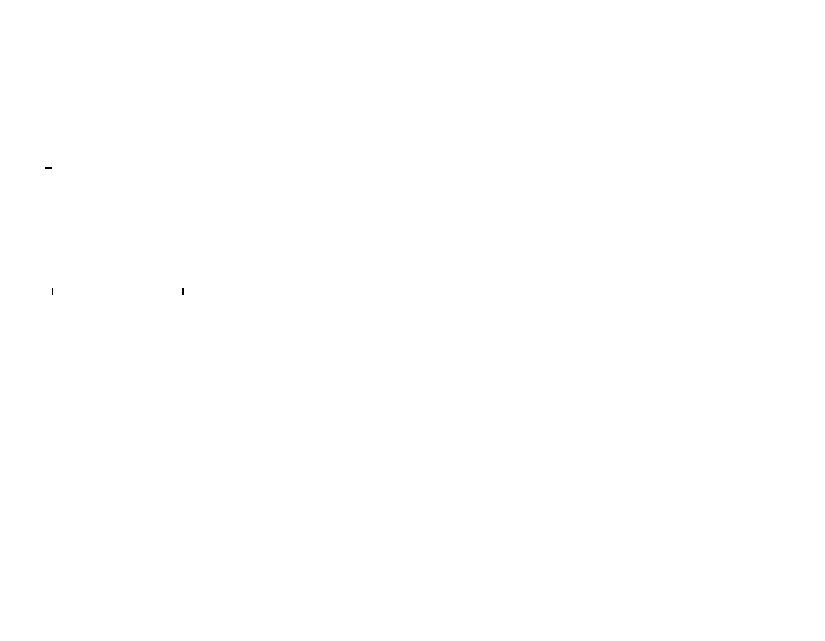\\
\end{minipage}
\begin{minipage}[b]{0.5\textwidth}
\end{minipage}

\caption[]{Comparison of $\psi$-ICNet to single branch ICNet. \textbf{a} and \textbf{b} show examples of real and predicted MUA in response to clean speech and music respectively. The left column shows real MUA recorded from three animals \textemdash one NH, one mild HI and one severe HI. The centre and right columns show MUA predicted by the single branch models and $\psi$-ICNet respectively.}
\label{fig:results_comparison}
\end{figure*}

Table \ref{tab:results} reports the median FEVE and KL divergence of the single branch ICNets on the test set, and the median pairwise difference to this score $\pm$ the median absolute deviation (MAD) over all sounds for the other models. We used the median and MAD since there were several outliers, mainly from the speech in noise and ripples sounds. The single branch model for a given animal generally achieves the best FEVE, but the 6 parameter $\psi$-ICNet comes very close for HI animals, and matches or outperforms the single branch model in terms of KL divergence. ABRNet performs worse than all other models for HI animals and worse than the 6 parameter $\psi$ model for NH animals. Figure \ref{fig:results_comparison} provides visual examples of the responses predicted by the $\psi$-ICNet model compared to the true responses and the predictions of the single branch models.

outloo\begin{table}
\centering
\fontsize{9pt}{9pt}\selectfont
\begin{tabularx}{\linewidth}{rllll} \hline
Model         & \multicolumn{2}{c}{FEVE}    &\multicolumn{2}{c}{KL / $10^{-3}$} \\
              &\multicolumn{1}{c}{NH}&\multicolumn{1}{c}{HI}&\multicolumn{1}{c}{NH}&\multicolumn{1}{c}{HI}\\
ICNet         &$67\pm14$ &$71\pm11$ &$8.4\pm2$ &$10.4\pm4$ \\ \hline
ABRNet        &$-5.6\pm4$&$-3.7\pm10$&$-0.36\pm3$&$-0.35\pm6$\\ 
$\psi$Net - 3&$-7.2\pm5$&$-3.5\pm10$&$-0.98\pm3$&$-1.0\pm4$\\
$\psi$Net - 6&$-5.3\pm5$&$-2.1\pm7$&$-1.1\pm3$&$-1.3\pm4$\\ \hline
\end{tabularx}
\caption{Median pairwise difference and median absolute deviation between the single branch model and $\psi$-ICNets, in terms of weighted FEVE and KL divergence. Each metric is computed over all sounds, split between NH and HI animals.}
\label{tab:results}
\end{table}

\graphicspath{{./figures/results_comparison}}

\section{Generating Data From Unseen Animals}
\label{sec:FittingtoUnseenAnimal}

We used a simple Bayesian optimisation routine to search over the conditional parameters $\psi$ (leaving $\theta$ fixed) of the trained models to find the optimal values $\psi^*$ for four animals (of which two were included in the training set, and two were not). We initialised the Gaussian process with 4, 8 or 16 evenly spaced random samples drawn from a Sobol sequence. At each step we evaluated the model's performance on a 30 minute subset of the training data consisting of 15 seconds from each sound in the training data which was collected in the first 2 hours of the recording, so as to keep the effects of time degradation to a minimum. The Bayesian optimisation routine was stopped when it found a candidate for $\psi^*$ which produced a loss within 2\% of the target loss. The target loss was taken as the value of the loss achieved using the learned means of the $\psi$ distributions. 

While we found that these models could converge quickly to a close-to-optimal $\psi$ for both in-sample and unseen animals, actual performance is significantly worse on unseen animals. This may simply be due to the low number of animals included in the space, so we also trained the 6 parameter model on data from 10 NH and 10 symmetric HI animals. We used the same hyperparameters except for the initial learning rate, which we set lower at $1\times10^{-4}$. We tested the 9 animal and 20 animal versions on 4 held out animals which were not seen by either model, and 4 in-sample animals which were included in training data for both models. We used a subset of the test data which was recorded in the first few hours of the recording for evaluation, including one 30 second sound from each sound class.

\subsection{Results}
\label{sec:Results}

\begin{figure*}
\centering 
\def\svgwidth{0.97\textwidth}
\fontsize{8pt}{8pt}\selectfont
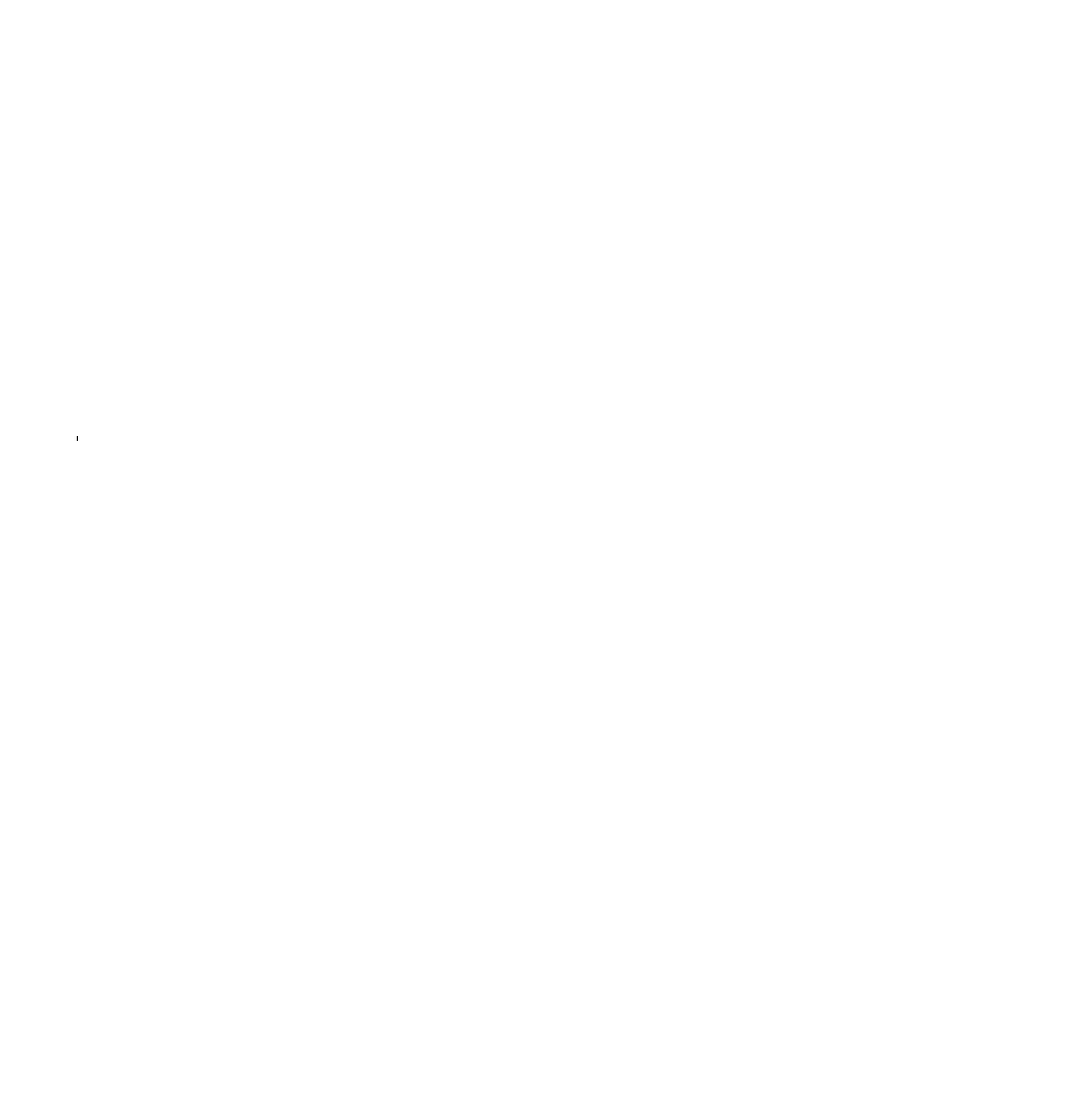

\caption[]{\textbf{a} and \textbf{b} show the number of iterations required to converge to the target loss when starting with 4, 8, or 16 random samples. \textbf{c} to \textbf{f} show the loss surface over a 2D projection of 1000 evenly spaced points in the $\psi$ space for 4 animals \textemdash two NH and two HI, two in-sample and two out of sample \textemdash for the 3 parameter $\psi$-ICNet. The colour of each point represents the loss obtained by generating MUA from the trained model with $\psi$ fixed to the given values. The black crosses represent the values of $\psi^*$ found by Bayesian optimisation over repeated runs, and the red star shows the $\psi$ value learned during training (or the best point found in a grid search for unseen animals). \textbf{g} to \textbf{j} show the same plots for the 6 parameter model, for which 4096 points in the $\psi$ space were evaluated in the grid search. \textbf{k} shows real (left) and simulated (right) MUA in response to a segment of speech for a model trained on 20 animals, 10 NH and 10 HI, when fitted to 4 different unseen animals. The hearing status of the animal is indicated on the right of the row. \textbf{l} and \textbf{m} show the relative FEVE of the 9 animal vs the 20 animal model for in-sample and out of sample animals, with NH animals highlighted in orange and HI animals in blue. \textbf{n} and \textbf{o} show the relative KL divergence. Different markers represent different sound classes. Key: R = ripples, SN = speech in noise, S = clean speech, M = music, I = single instruments.}
\label{fig:results_bayesopt}
\end{figure*}

For the 3 parameter model, Bayesian optimisation generally converged to a $\psi^*$ which achieved the target loss for in-sample animals within 10-20 iterations. The 6 parameter model required 15-35 iterations, shown in figures \ref{fig:results_bayesopt}\textbf{a} and \ref{fig:results_bayesopt}\textbf{b}. 

Figures \ref{fig:results_bayesopt}\textbf{c} to \textbf{j} show the results of our Bayesian optimisation over the conditioning parameters $\psi$. The optimal $\psi^*$ values found by Bayesian optimisation (the black crosses) are clustered around the learned value of $\psi$ for each animal (the red stars), even for the severely hearing impaired animal in figure \ref{fig:results_bayesopt}\textbf{e} for which only a small area of the $\psi$ space produces a good loss. The 3 parameter model learns $\psi$ parameters with a smoothly varying loss surface for both HI and NH animals, in and out of sample. In particular, figures \ref{fig:results_bayesopt}\textbf{c} and 
\ref{fig:results_bayesopt}\textbf{d}, representing in/out of sample NH animals, have very similar loss surfaces. There is also a clear contrast between these normal hearing loss surfaces and the loss surfaces for the HI animals shown in figures \ref{fig:results_bayesopt}\textbf{e} and \ref{fig:results_bayesopt}\textbf{f}. Figures \ref{fig:results_bayesopt}\textbf{g} to \textbf{j} suggest that the same holds even for models with six $\psi$ parameters, although the results are harder to interpret since we have projected six dimensions on to two. 

Finally, figure \ref{fig:results_bayesopt}\textbf{k} shows real and simulated MUA for 4 unseen animals with varying hearing status generated by the trained model using $\psi^*$ found by Bayesian optimisation. From figures \ref{fig:results_bayesopt}\textbf{l} to \textbf{o}, we see that training the 6 parameter model with more animals improved the median FEVE to 26.9\% from 7.8\% on unseen animals, while degrading the median FEVE on in-sample animals from 55.6\% to 50.7\%.

\graphicspath{{./figures/results_bayesopt}}

\section{Discussion and Limitations}
\label{sec:Discussion}

Although we demonstrated that we are able to learn a small set of control parameters which lead to a smoothly varying expression of hearing loss over a range of animals, the single branch ICNet architecture is still slightly better at predicting the MUA of a specific animal, in terms of FEVE, over most sounds. When fitting the $\psi$ parameters to unseen animals, performance is worse than an animal specific model trained from scratch. This is likely in part due to unseen animals being disproportionately affected by residual misalignment, which can be fixed to some degree by the conditioning parameters but which does not generalise to unseen animals.

Increasing the number of animals in the training dataset improved the performance of the model on unseen animals, but degraded the performance on in-sample animals, as shown in figures \ref{fig:results_bayesopt}\textbf{l} and \textbf{n}. This could indicate that the model is under-parametrised to fully capture hearing loss, or that the model is learning to correct less non-hearing-related idiosyncrasies (i.e. alignment), leading to a convergence between the performance on seen and unseen animals.

\section{Conclusion}
\label{sec:Conclusion}

We have described a novel variational-conditional model of neural activity in the auditory midbrain which models hearing loss by learning variational conditioning parameters directly from the data. We have shown that this model performs comparably to existing DNN models trained from scratch on specific brains with varying hearing ability, and that by learning conditioning parameters directly from the data we achieve better performance than an equivalent architecture conditioned on audiometric thresholds. 

We also demonstrated that we can quickly (within 15-30 iterations) optimise the model for a new brain, fitting only the conditioning parameters $\psi$ with Bayesian optimisation, and that the space of $\psi$ parameters encodes a smoothly varying spectrum of hearing loss. We showed that by fitting the $\psi$ parameters we can generate MUA for unseen animals which qualitatively resembles real recorded MUA, and which accurately predicts 25+\% of the explainable variance in the neural responses. Although this is far off the performance on in-sample animals, we also showed that performance increases as we add more animals to the training dataset.

\section{Future Work and Applications}
\label{sec:FutureWork}

Core sound processing in current hearing aids is limited to frequency dependent amplification and compression, but we know that the real effects of hearing loss are more complex than loss of sensitivity to certain frequencies and shifted dynamic range \cite{lesica2018hearing, parida2022underlying}. Amplification may restore audibility in situations where relevant audio stimuli are too quiet to hear, but it cannot restore lost intelligibility caused by more complex distortions of the neural code. One approach to developing hearing aids beyond the current state of the art is to use auditory models of NH and HI brains to develop sound processing algorithms which directly correct the distorted neural code \cite{bondy2004novel}.

Recent work has explored training DNN-based hearing compensation strategies on the outputs of DNN-based models of the NH and HI auditory system \cite{drakopoulos2023neural, leer2025hearing}. This has recently led to the first subjective evaluations of deep-learning based hearing compensation in listening tests with hearing impaired listeners \cite{leer2025hearing}, in which DNN-based hearing compensation outperforms state of the art hearing aids.

Our model is uniquely well suited for this application since we simulate activity in the midbrain rather than the ear, and we train directly on recorded data. A parametrised hearing compensation model could be trained to correct distortions in the neural code for a wide spectrum of hearing loss by sampling from the learned space to generate realistic activity. This would allow development of a hearing aid with a small set of control parameters directly learned from the effects of hearing loss on the neural code, which could be quickly fitted to a new listener by human in the loop optimisation. The success of our model in generating neural activity for brains with unseen hearing loss profiles by searching over the $\psi$ space gives us confidence in this approach.

\section*{Acknowledgements}
This work was supported by EPSRC EP/W004275/1 and BBSRC BB/Y008758/1. We also thank Prof. Maneesh Sahani for his helpful feedback

\bibliography{aaai2026}

\end{document}